\documentclass[a4paper]{jpconf}
\usepackage{graphicx}
\usepackage{amssymb}
\usepackage{amsmath}
\usepackage{fancyhdr}
\usepackage{hyperref}
\renewcommand{\thispagestyle}[1]{} % do nothing
 \def\be{\begin{equation}}
 \def\ee{\end{equation}}
 \def\bea{\begin{eqnarray}}
 \def\eea{\end{eqnarray}}
 \def\bean{\begin{eqnarray*}}
 \def\eean{\end{eqnarray*}}

\begin{document}
\title{Nonequilibrium dynamics and transport near the chiral phase transition of a quark-meson model}

\author{A Meistrenko$^1$, C Wesp$^1$, H van Hees$^{1,2}$ and C Greiner$^1$ }

\address{$^1$ Institut f\"ur Theoretische Physik, Goethe-Universit\"at Frankfurt, Max-von-Laue-Stra{\ss}e 1, D-60438 Frankfurt, Germany}
\address{$^2$ Frankfurt Institute for Advanced Studies, Ruth-Moufang-Stra{\ss}e 1, D-60438 Frankfurt, Germany}

\ead{meistrenko@th.physik.uni-frankfurt.de}

\begin{abstract}
Based on the 2PI quantum effective action of the linear sigma model with constituent quarks, we develop a transport approach to study systems out of equilibrium. In particular, we focus on the chiral phase transition as well as the critical point, where nonequilibrium effects near the phase transition give rise to critical behavior such as the fluctuation of the baryon number density. Predictions for long-range correlations and fluctuations of observables in our model could be used to study fundamental properties of the QCD phase transition. In the previous version of our transport model the chiral fields are implemented as mean fields, whereas quarks are treated as on-shell particles in the Vlasov equation with a dynamical force term. The current update includes also the distribution functions of sigma mesons and pions in a self-consistent way. On this basis a dissipation kernel between the mean fields and particle modes can be implemented.
\end{abstract}

\section{Introduction}
\label{sec:a1}
We concentrate our study on nonequilibrium effects and the chiral phase transition within the linear sigma model with constituent quarks, which is an effective theory of QCD in the low-energy limit. Our motivation is to investigate critical phenomena at the phase transition and the critical point, where time-dependent long-range correlations can arise. Such effects can be studied in a (3+1)-dimensional numerical approach. We present our model in the following section \ref{sec:a2} and discuss an improved set of transport equations in section \ref{sec:a3}. Finally, we summarize the progress in section \ref{sec:a4} and give an outlook to further improvements.

\section{Classical transport equations within a linear sigma framework}
\label{sec:a2}
The linear sigma model with constituent quarks is a well known $O\left(4\right)$-model \cite{Scavenius:2000qd}, which is suited for studying the chiral phase transition. Because of the spontaneously and explicitly broken chiral symmetry the mesonic part of this theory consists of a massive scalar sigma field and three isoscalar pion fields, which form the chiral field $\Phi=\left(\sigma, \vec\pi\right)$. The sigma field represents the order parameter for the chiral phase transition and mimics the properties of the quark condensate in QCD, since both transform equally under chiral transformation. Without explicit symmetry breaking a $\mathrm{SU_L}\left(2\right)\times \mathrm{SU_R}\left(2\right)$ symmetry transformation would let the Lagrangian invariant. It has the following form:
 \be
 \begin{aligned}
 \mathcal L=&\sum_i\bar\psi_i \Big[i\partial\!\!\!/-g\left(\sigma+i\gamma_5\vec\pi\cdot\vec\tau\right)\Big]\psi_i
 +\frac{1}{2}\left(\partial_\mu\sigma\partial^\mu\sigma
 +\partial_\mu\vec\pi\partial^\mu\vec\pi\right)\\
 &-\frac{\lambda}{4}\left(\sigma^2+\vec\pi^2-\nu^2\right)^2+f_{\pi}m_\pi^2\sigma+U_0\,,
 \end{aligned}
 \label{eqn:a1}
 \ee
where the field shift term and the zero potential constant are given by $\nu^2=f_\pi^2-m_\pi^2/\lambda$, $U_0=m^4_\pi/\left(4\lambda\right)-f_\pi^2m_\pi^2$. The sum runs over included quark flavor $\psi_i$ and the parameters of the model are adjusted to match the vacuum values of the pion decay constant $f_\pi=93\,\mbox{MeV}$, the pion mass $m_\pi=138\,\mbox{MeV}$ as well as an estimated sigma mass $m_\sigma\approx604\,\mbox{MeV}$.

\begin{figure}[h]
\begin{minipage}{18pc}
\includegraphics[width=18pc]{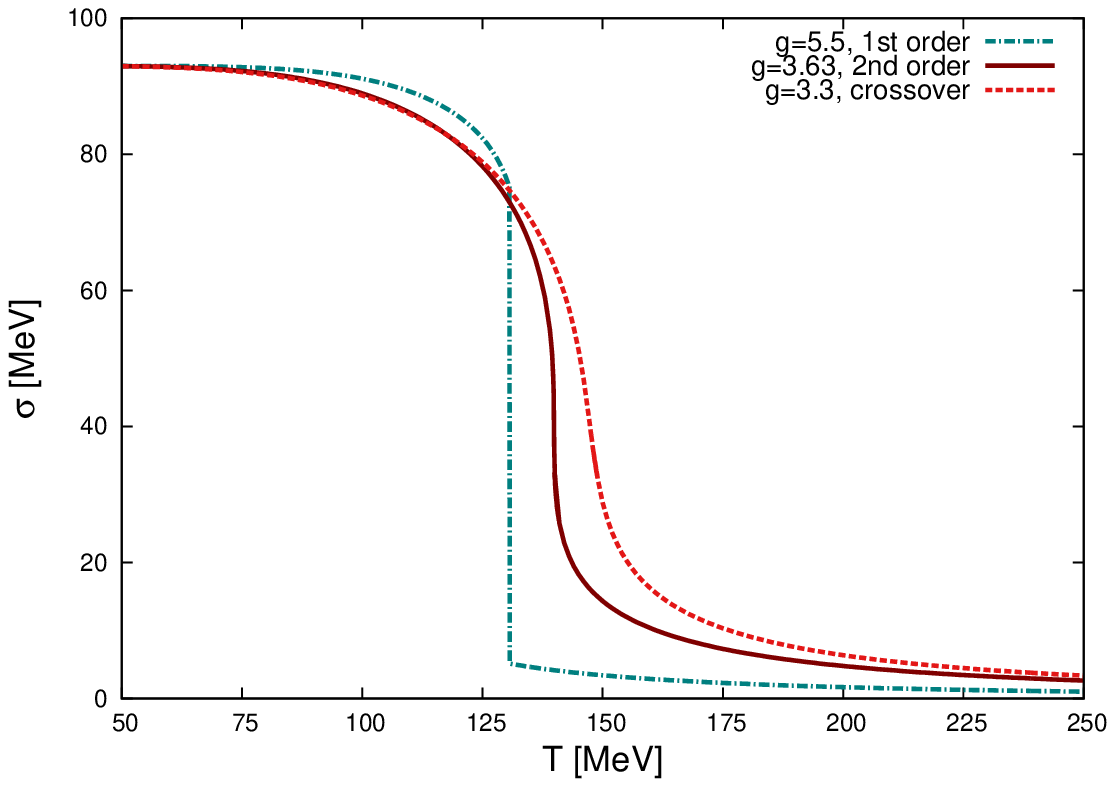}
\caption{\label{fig:a1}Order parameter sigma as a function of the temperature $T$ for $3$ different values of the Yukawa coupling $g$.}
\end{minipage}\hspace{2pc}%
\begin{minipage}{18pc}
\vspace{-0.2pc}
\includegraphics[width=18pc]{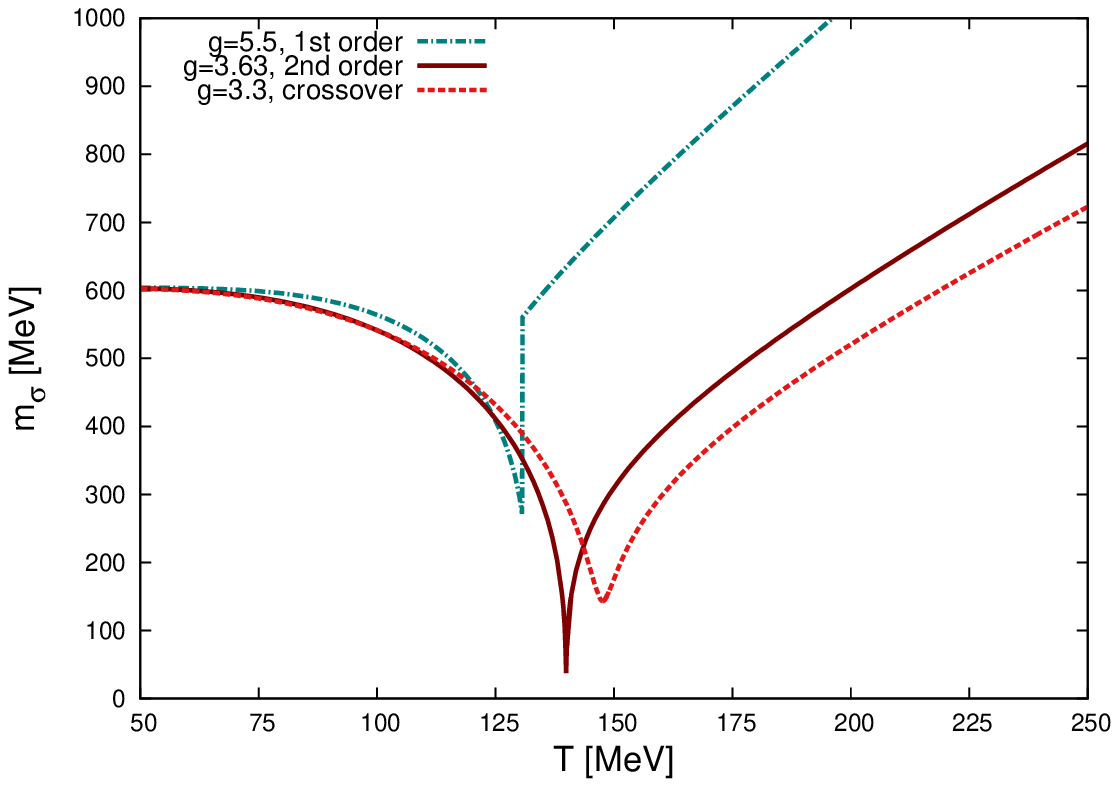}
\caption{\label{fig:a2}Same as Fig. \ref{fig:a1} but for the sigma mass spectrum.\newline}
\end{minipage}
\end{figure}

The inhomogeneous Klein-Gordon equations of motion for the mean fields follow directly from the functional derivative of the classical action with respect to the chiral field components and treating quarks at one-loop level
\be
\begin{aligned}
\partial_\mu\partial^\mu\sigma+\lambda\left(\sigma^2+\vec\pi^2-\nu^2\right)\sigma-f_\pi m_\pi^2+g\left<\bar\psi\psi\right>&=0\,,\\
\partial_\mu\partial^\mu\vec\pi+\lambda\left(\sigma^2+\vec\pi^2-\nu^2\right)\vec\pi+g\left<\bar\psi i\gamma_5\vec\tau\psi\right>&=0\,.
\end{aligned}
\label{eqn:a2}
\ee
The scalar and pseudoscalar densities are given by
\be
\begin{aligned}
\left<\bar\psi\psi\right>\left(t,\vec x\right)&=gd_q\sigma\left(t,\vec x\right)\int d^3\vec p\,\frac{f_q\left(t,\vec x,\vec p\right)+f_{\bar q}\left(t,\vec x,\vec p\right)}{E\left(t,\vec x,\vec p\right)}\,,\\
\left<\bar\psi i\gamma_5\vec\tau\psi\right>\left(t,\vec x\right)&=gd_q\vec\pi\left(t,\vec x\right)\int d^3\vec p\,\frac{f_q\left(t,\vec x,\vec p\right)+f_{\bar q}\left(t,\vec x,\vec p\right)}{E\left(t,\vec x,\vec p\right)}\,,
\end{aligned}
\label{eqn:a3a}
\ee
where $f_q$ and $f_{\bar q}$ denote the phase-space distribution functions of quarks and antiquarks with their degeneracy factor $d_q$. Numerically the quarks are treated as test particles, which propagate according to the Vlasov equation
\be
\left[\partial_t+\frac{p}{E\left(t,\vec x,\vec p\right)}\cdot\nabla_{\vec x}-\nabla_{\vec x} E\left(t,\vec x,\vec p\right)\nabla_{\vec p}\right]f_{q|\bar q}\left(t,\vec x,\vec p\right)=0\,.
\label{eqn:a3b}
\ee
Thereby the force term is dynamically generated through the effective mass of quarks, depending on the mean-field values
\be
\begin{aligned}
E\left(t,\vec x,\vec p\right)&=\sqrt{\vec{p}^2\left(t\right)+M^2\left(t,\vec x\right)}\,,\\
M^2\left(t,\vec x\right)&=g^2\left[\sigma^2\left(t,\vec x\right)+\vec\pi^2\left(t,\vec x\right)\right]\,.
\end{aligned}
\label{eqn:a3c}
\ee
The resulting equilibrium properties of such a classical system are shown in Fig. \ref{fig:a1} for the order parameter and in Fig. \ref{fig:a2} for the effective sigma mass as a function of the temperature. Depending on the Yukawa coupling constant $g$ different order of a phase transition can be generated. More results, also with a constant binary cross section for quarks, can be found in \cite{H}.

\section{Transport equations from the 2PI effective action}
\label{sec:a3}
Since we are interested in nonequilibrium dynamics at the phase transition, a consistent real-time description of the interaction between the mean-field (soft) and the particle (hard) modes of the chiral field as well as quarks is needed. The model should also include a dissipation term, which would drive the system to thermal equilibrium. In the following we use the 2PI quantum effective action to derive a set of coupled equations of motion. Thereby 2PI stands for two-particle irreducible, which means that a 2PI diagram does not become disconnected by cutting $2$ inner lines (propagators). This effective action for the mean fields and propagators preserves global symmetries of the original theory, guarantees thermodynamic consistency \cite{Baym:1962} and can be renormalized with vacuum counter terms \cite{vanHees:2001ik,VanHees:2001pf}. At the same time such a self-consistent approximation is suitable to derive off-shell transport equations, which also respect conservation laws and the correct equilibrium limit. Furthermore, a systematic inclusion of collisional memory effects is possible \cite{Ivanov:1998nv, Ivanov:1999tj, Knoll:2001jx, Ivanov:2003wa}.

\begin{figure}[h]
\includegraphics[width=28pc]{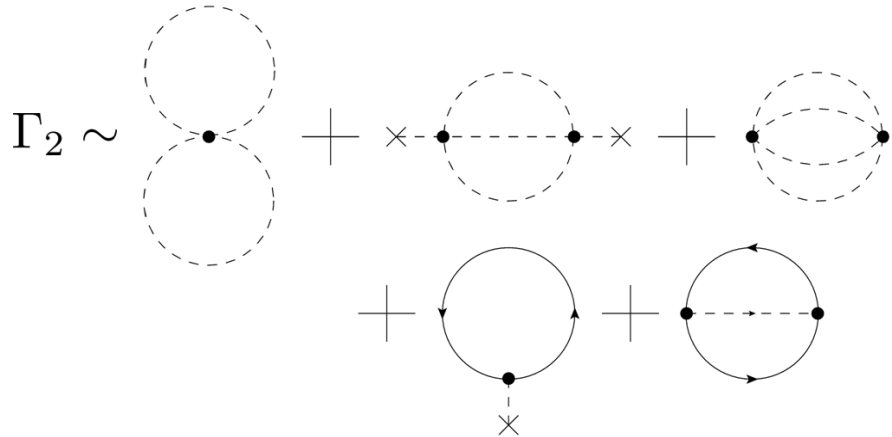}
\centering
\caption{Approximation for the 2PI part of the effective action: the upper line shows bosonic diagrams (Hartree, sunset and basketball), where dashed lines stand for boson propagators and external lines with a cross represent mean fields. The lower line shows interactions between bosons and fermions, here a solid line stands for a fermion propagator.}
\label{fig:a3}
\end{figure}

A more phenomenological study, motivated by the Langevin equation, is discussed in \cite{H}, where a new statistical approach for the scattering effects between quarks as quasi particles and mean fields is introduced. Within the linear sigma model with constituent quarks \eqref{eqn:a1} the 2PI effective action reads
\be
\begin{aligned}
\Gamma[\sigma,\vec\pi,G,D] = & S_{cl}[\sigma,\vec\pi]+\frac{i}{2}\cdot\mbox{Tr}\log G^{-1}+\frac{i}{2}\cdot\mbox{Tr}\,G_0^{-1}G\\
 & -i\cdot\mbox{Tr}\log D^{-1}-i\cdot\mbox{Tr}\,D_0^{-1}D+\Gamma_2[\phi,\vec\pi,G,D]\,,
\end{aligned}
\label{eqn:a4}
\ee
where $G_0,\,D_0$ denote the free and $G,\,D$ the fully dressed propagators for bosons and fermions, which fulfills the Schwinger-Dyson equation and are formally given by
\be
\begin{aligned}
G^{-1}\left(x,y\right)&=G_0^{-1}\left(x,y\right)-\Pi\left(x,y\right)\,,\qquad
\Pi\left(x,y\right)=2i\frac{\delta\Gamma_2}{\delta G\left(x,y\right)}\,,\\
D^{-1}\left(x,y\right)&=D_0^{-1}\left(x,y\right)-\Sigma\left(x,y\right)\,,\qquad
\Sigma\left(x,y\right)=-i\frac{\delta\Gamma_2}{\delta D\left(y,x\right)}\,.
\end{aligned}
\label{eqn:a5}
\ee
Here, $\Pi$ and $\Sigma$ denote the bosonic and fermionic self-energies. These are derived from the 2PI part of the effective action, where we include only $1$- and $2$-point diagrams as shown in Fig. \ref{fig:a3}, since a first order gradient expansion of Kadanoff-Baym and Schwinger-Dyson equations in Wigner space then reduces to a Markov-like collisional dynamics without memory effects \cite{Ivanov:1998nv, Ivanov:1999tj, Knoll:2001jx, Ivanov:2003wa}. Unfortunately, finite truncations of the 2PI effective action violate Ward-Takahashi-identities of global and local symmetries in the first neglected order of the expansion parameter, that is a direct consequence of the resummation in the $2$-point function. It follows, that also the Goldstone theorem is violated \cite{Baym:1977,vanHees:2002bv}, and the pions acquire a non-vanishing (temperature dependent) mass in the broken phase of the linear sigma model, even without explicitly broken chiral symmetry. Possible modifications for a symmetry-improved 2PI effective action are discussed in \cite{Pilaftsis:2013xna}.

In a first attempt we extend our classical set of equations \eqref{eqn:a2}, \eqref{eqn:a3b} by including also the Vlasov equation for the phase-space distribution functions of $\sigma$ and $\pi$ mesons in a Hartree-like approximation, where only the local part of $\Gamma_2$ is considered. Such a procedure requires the computation of self-consistently derived effective mass terms, which follow from the gap equation of the propagator. Starting from the general expression for the bosonic propagator \eqref{eqn:a5} and inserting the self-energy expressions from the Hartree diagrams, leads to the following self-consistent equations
\be
\begin{aligned}
M_\sigma^2\left(x\right)&=\lambda\left(3\sigma^2+3\pi^2-\nu^2\right)+3\lambda G_{\sigma\sigma}+3\lambda G_{\pi\pi}\,,\\
M_\pi^2\left(x\right)&=\lambda\left(\sigma^2+5\pi^2-\nu^2\right)+\lambda G_{\sigma\sigma}+5\lambda G_{\pi\pi}\,,
\end{aligned}
\label{eqn:a6}
\ee
where the bosonic loop integrals are given by
\be
G_{\phi\phi}\left(t,\vec x\right)=\frac{1}{2}\int\frac{d^3p}{\left(2\pi\right)^2}\frac{1+2N_\phi\left(t,\vec x,\vec p\right)}{\sqrt{\vec p^2+M^2_\phi\left(t,\vec x\right)}}\quad\mbox{with}\quad\phi\in\{\sigma,\pi\}\,.
\label{eqn:a7}
\ee
The Vlasov equations for the phase-space distributions of $\sigma$ and $\pi$ mesons are then given by
\be
\left[\partial_t+\frac{p}{E_\phi\left(t,\vec x,\vec p\right)}\cdot\nabla_{\vec x}-\nabla_{\vec x} E_\phi\left(t,\vec x,\vec p\right)\nabla_{\vec p}\right]f_\phi\left(t,\vec x,\vec p\right)=0\,,
\label{eqn:a8}
\ee
where the force term depends on the self-consistent mass expressions from \eqref{eqn:a6} as gradient of
\be
E_\phi\left(t,\vec x,\vec p\right)=\sqrt{\vec{p}^2+M_\phi^2\left(t,\vec x\right)}\,.
\label{eqn:a9}
\ee
The Vlasov equations of the chiral field components are solved as differential equations without using a test particle ansatz.
Including also the tadpole contribution of $\Gamma$, we end up with the following mean-field equations
\be
\begin{aligned}
\partial_\mu\partial^\mu\sigma+\lambda\left(\sigma^2+\vec\pi^2-\nu^2+3G_{\sigma\sigma}+3G_{\pi\pi}\right)\sigma-f_\pi m_\pi^2+g\left<\bar\psi\psi\right>&=0\,,\\
\partial_\mu\partial^\mu\vec\pi+\lambda\left(\sigma^2+\vec\pi^2-\nu^2+G_{\sigma\sigma}+5G_{\pi\pi}\right)\vec\pi+g\left<\bar\psi i\gamma_5\vec\tau\psi\right>&=0\,.
\end{aligned}
\label{eqn:a10}
\ee
The system of equations \eqref{eqn:a3b}, \eqref{eqn:a8} and \eqref{eqn:a10} serves as a basis to study nonequilibrium effects. Nevertheless, it does not account for dissipation as well as $q,\bar q$ creation and annihilation processes. In an upcoming work we will implement a dissipation kernel \cite{Greiner:1996dx,Rischke:1998qy}, which arises from the sunset diagram (see Fig. \ref{fig:a3}) by deriving $\Gamma_2$ with respect to the mean fields.

\section{Conclusion and outlook}
\label{sec:a4}
In this proceeding we presented an improved set of equations of motion to study systems out of equilibrium, which is included in a numerical simulation. In comparison to the phenomenological work \cite{H}, here we use the 2PI effective action of the linear sigma model with constituent quarks. In the current version the set consists of mean-field equations for the components of the chiral field, the Vlasov equation for the phase-space distribution functions of $\sigma$ and $\pi$ mesons as well as the Vlasov equation for the quarks, which are treated as test particles.

In a further study we plan to implement a dissipation kernel for the mean-field equations, which results from the sunset diagram. For a consistent treatment it requires also the inclusion of a collision term between soft and hard modes for the Vlasov equation of chiral partners. Furthermore, an interaction between $\sigma$, $\pi$ and quarks on the particle level has to be implemented to account for chemical equilibration in a dynamical and consistent way. Because of numerical challenges such a direct Boltzmann-like transport approach will be primarily suited for the study of time-dependent fluctuations in homogeneous systems. High order cumulants of conserved quantities like baryon number density offer a possibility to observe such fluctuations. Nevertheless, it is an open question whether fluctuations can also arise on the time scale of a heavy ion collision and survive the hadronization, that needs further investigation.

\ack
This work has been supported by the German Federal Ministry of Education and Research (BMBF F\"orderkennzeichen 05P12RFFTS). A. M. and C. W. acknowledge financial support from the Helmholtz Research School for Quark Matter Studies (H-QM) and HIC for FAIR.


\section*{References}
\begin{thebibliography}{9}
%\cite{Scavenius:2000qd}
\bibitem{Scavenius:2000qd}
  Scavenius O, Mocsy A, Mishustin I N and Rischke D H
  %``Chiral phase transition within effective models with constituent quarks,''
  2001 {\it Phys. Rev.} C {\bf 64}, 045202
  %%CITATION = NUCL-TH/0007030;%%
  %200 citations counted in INSPIRE as of 24 Nov 2013
\bibitem{H}
    van Hees H, Wesp C, Meistrenko A and Greiner C 2013
    Dynamics of the chiral phase transition
    {\it Preprint to be published in the proceedings of the XXXI Max Born Symposium and HIC for FAIR Workshop, 14-16 June 2013, Wroclaw, Poland}
\bibitem{Baym:1962}
  Baym G
  1962 {\it Phys. Rev.} {\bf 127}, 1391
  %\cite{vanHees:2001ik}
\bibitem{vanHees:2001ik}
  van Hees H and Knoll J
  %``Renormalization in selfconsistent approximations schemes at finite temperature. 1. Theory,''
  2002 {\it Phys. Rev.} D {\bf 65}, 025010
  %[hep-ph/0107200].
  %%CITATION = HEP-PH/0107200;%%
  %129 citations counted in INSPIRE as of 24 Nov 2013
  %\cite{VanHees:2001pf}
\bibitem{VanHees:2001pf}
  van Hees H and Knoll J
  %``Renormalization of selfconsistent approximation schemes. 2. Applications to the sunset diagram,''
  2002 {\it Phys. Rev.} D {\bf 65}, 105005
  %[hep-ph/0111193].
  %%CITATION = HEP-PH/0111193;%%
  %114 citations counted in INSPIRE as of 24 Nov 2013
  %\cite{Ivanov:1998nv}
\bibitem{Ivanov:1998nv}
  Ivanov Y B, Knoll J and Voskresensky D N
  %``Selfconsistent approximations to nonequilibrium many body theory,''
  1999 {\it Nucl. Phys.} A {\bf 657}, 413
  %[hep-ph/9807351].
  %%CITATION = HEP-PH/9807351;%%
  %76 citations counted in INSPIRE as of 24 Nov 2013
  %\cite{Ivanov:1999tj}
\bibitem{Ivanov:1999tj}
  Ivanov Y B, Knoll J and Voskresensky D N
  %``Resonance transport and kinetic entropy,''
  2000 {\it Nucl. Phys.} A {\bf 672}, 313
  %[nucl-th/9905028].
  %%CITATION = NUCL-TH/9905028;%%
  %83 citations counted in INSPIRE as of 24 Nov 2013
  %\cite{Knoll:2001jx}
\bibitem{Knoll:2001jx}
  Knoll J, Ivanov Y B and Voskresensky D N
  %``Exact conservation laws of the gradient expanded Kadanoff-Baym equations,''
  2001 {\it Annals Phys.}  {\bf 293}, 126
  %[nucl-th/0102044].
  %%CITATION = NUCL-TH/0102044;%%
  %43 citations counted in INSPIRE as of 24 Nov 2013
  %\cite{Ivanov:2003wa}
\bibitem{Ivanov:2003wa}
  Ivanov Y B, Knoll J and Voskresensky D N
  %``Selfconsistent approach to off-shell transport,''
  2003 {\it Phys. Atom. Nucl.}  {\bf 66}, 1902
  %[nucl-th/0303006].
  %%CITATION = NUCL-TH/0303006;%%
  %19 citations counted in INSPIRE as of 24 Nov 2013
%\cite{Arrizabalaga:2002hn}
\bibitem{Arrizabalaga:2002hn}
  Arrizabalaga A and Smit J
  %``Gauge fixing dependence of Phi derivable approximations,''
  2002 {\it Phys. Rev.} D {\bf 66}, 065014
  %[hep-ph/0207044].
  %%CITATION = HEP-PH/0207044;%%
  %81 citations counted in INSPIRE as of 24 Nov 2013
\bibitem{Baym:1977}
  Baym G and Grinstein G
  1977 {\it Phys. Rev.} D {\bf 15}, 2897–2912
  %\cite{vanHees:2002bv}
\bibitem{vanHees:2002bv}
  van Hees H and Knoll J
  %``Renormalization in selfconsistent approximation schemes at finite temperature. 3. Global symmetries,''
  2002 {\it Phys. Rev.} D {\bf 66}, 025028
  %[hep-ph/0203008].
  %%CITATION = HEP-PH/0203008;%%
  %99 citations counted in INSPIRE as of 24 Nov 2013
  %\cite{Pilaftsis:2013xna}
\bibitem{Pilaftsis:2013xna}
  Pilaftsis A and Teresi D
  %``Symmetry Improved CJT Effective Action,''
  2013 {\it Nucl. Phys.} B {\bf 874}, 594
  %[arXiv:1305.3221 [hep-ph]].
  %%CITATION = ARXIV:1305.3221;%%
  %2 citations counted in INSPIRE as of 24 Nov 2013
%\cite{Greiner:1996dx}
\bibitem{Greiner:1996dx}
  Greiner C and Muller B
  %``Classical fields near thermal equilibrium,''
  1997 {\it Phys. Rev.} D {\bf 55}, 1026
  %[hep-th/9605048].
  %%CITATION = HEP-TH/9605048;%%
  %102 citations counted in INSPIRE as of 24 Nov 2013
  %\cite{Rischke:1998qy}
\bibitem{Rischke:1998qy}
  Rischke D H
  %``Forming disoriented chiral condensates through fluctuations,''
  1998 {\it Phys. Rev.} C {\bf 58}, 2331
  %[nucl-th/9806045].
  %%CITATION = NUCL-TH/9806045;%%
  %53 citations counted in INSPIRE as of 24 Nov 2013
  \end{thebibliography}
\end{document}